\documentclass[prd,twocolumn,nofootinbib,superscriptaddress,amssymb,amsfonts,amsmath,preprintnumbers]{revtex4-2}

\DeclareUnicodeCharacter{2212}{\ensuremath{-}}

\usepackage{graphicx,slashed,xcolor,multirow,bbold,mathtools,sidecap,tikz,bm,ulem,enumitem,booktabs,array}
\usepackage[colorlinks=true,linktoc=page,linkcolor=purple,citecolor=teal,urlcolor=magenta]{hyperref}

\definecolor{lime}{HTML}{A6CE39}
\DeclareRobustCommand{\orcidicon}{\hspace{-2.1mm}
\begin{tikzpicture}
\draw[lime,fill=lime] (0,0.0) circle [radius=0.13] node[white] {{\fontfamily{qag}\selectfont \tiny ID}}; \draw[white,fill=white] (-0.0525,0.095) circle [radius=0.007]; 
\end{tikzpicture} \hspace{-3.7mm} }
\foreach \x in {A, ..., Z} {\expandafter\xdef\csname orcid\x\endcsname{\noexpand\href{https://orcid.org/\csname orcidauthor\x\endcsname} {\noexpand\orcidicon}}}

\usepackage{siunitx,adjustbox}

\begin{document}

\preprint{PSI-PR-23-21, ZU-TH 31/23, ICPP-70}

\title{Growing Excesses of New Scalars at the Electroweak Scale}

\author{Srimoy Bhattacharya\orcidB}
\email{bhattacharyasrimoy@gmail.com}
\affiliation{School of Physics and Institute for Collider Particle Physics, University of the Witwatersrand,
Johannesburg, Wits 2050, South Africa}

\author{Guglielmo Coloretti\orcidD{}}
\email{guglielmo.coloretti@physik.uzh.ch}
\affiliation{Physik-Institut, Universität Zürich, Winterthurerstrasse 190, CH–8057 Zürich, Switzerland}
\affiliation{Paul Scherrer Institut, CH–5232 Villigen PSI, Switzerland}

\author{Andreas Crivellin\orcidE{}}
\email{andreas.crivellin@psi.ch}
\affiliation{Physik-Institut, Universität Zürich, Winterthurerstrasse 190, CH–8057 Zürich, Switzerland}
\affiliation{Paul Scherrer Institut, CH–5232 Villigen PSI, Switzerland}

\author{Salah-Eddine Dahbi\orcidS{}}
\email{salah-eddine.dahbi@cern.ch}
\affiliation{School of Physics and Institute for Collider Particle Physics, University of the Witwatersrand, Johannesburg, Wits 2050, South Africa}

\author{Yaquan Fang}
\email{fangyq@ihep.ac.cn}
\affiliation{Institute of High Energy Physics, 19B, Yuquan Road, Shijingshan District, Beijing, 100049, China}
\affiliation{University of Chinese Academy of Sciences (CAS), 19A Yuquan Road, Shijingshan District, Beijing, 100049, China}

\author{Mukesh Kumar\orcidM{}}
\email{mukesh.kumar@cern.ch}
\affiliation{School of Physics and Institute for Collider Particle Physics, University of the Witwatersrand,
Johannesburg, Wits 2050, South Africa}

\author{Bruce Mellado}
\email{bmellado@mail.cern.ch}
\affiliation{School of Physics and Institute for Collider Particle Physics, University of the Witwatersrand,
Johannesburg, Wits 2050, South Africa}
\affiliation{iThemba LABS, National Research Foundation, PO Box 722, Somerset West 7129, South Africa}

\begin{abstract}
We combine searches for scalar resonances at the electroweak scale performed by the Large Hadron Collider experiments ATLAS and CMS where persisted excesses have been observed in recent years. Using both the side-bands of Standard Model Higgs analyses as well as dedicated beyond the Standard Model analyses, we find significant hints for new scalars at $\approx 95\,$GeV ($S^\prime$) and $\approx152\,$GeV ($S$). The presence of a $95\,$GeV scalar is preferred over the Standard Model hypothesis by $3.8\sigma$, while interpreting the $152\,$GeV excesses in a simplified model with resonant pair production of $S$ via a new heavier scalar $H(270)$, a global significance of $\approx5\sigma$ is obtained. While the production mechanism of the $S^\prime$ cannot yet be determined, data strongly favours the associated production of $S$, i.e.~via the decay of a heavier boson $H$ ($pp\to H\to SS^*$). A possible alternative or complementary decay chain is $H\rightarrow SS^{\prime}$, where $S\to WW^*$ ($S^{\prime}$) would be the source of the leptons ($b$-quarks) necessary to explain the multi-lepton anomalies found in Large Hadron Collider data. 
\end{abstract}

\maketitle

\section{Introduction}
The Standard Model (SM) of particle physics is the mathematical description of the fundamental constituents of matter and their interactions at microscopic scales. It has been extensively and successfully tested~\cite{ParticleDataGroup:2022pth,HeavyFlavorAveragingGroup:2022wzx,ALEPH:2005ab}, with the discovery of the Brout-Englert-Higgs boson ($h$)~\cite{Higgs:1964ia,Englert:1964et,Higgs:1964pj,Guralnik:1964eu} in 2012 at the Large Hadron Collider (LHC)~\cite{Aad:2012tfa,Chatrchyan:2012ufa} at CERN providing the last missing particle. Furthermore, measurements of the properties of this $125\,$GeV boson agree with the SM predictions~\cite{Chatrchyan:2012jja,Aad:2013xqa,Langford:2021osp,ATLAS:2021vrm}. 

Despite the overwhelming success of the SM, the existence of additional scalar bosons is not excluded as long as their role in electroweak symmetry breaking is sufficiently small. In fact, it is clear that the SM cannot be the ultimate fundamental theory of nature. It can neither account for the observed non-vanishing neutrino masses nor the existence of Dark Matter (DM) established at astrophysical scales. Moreover, the minimality of the SM Higgs sector, i.e.~the presence of a single $SU(2)_L$ doublet scalar that simultaneously gives mass to the electroweak (EW) gauge bosons and all fermions, is not guaranteed by any symmetry or principle. In fact most New Physics (NP) models in the literature contain new scalars, such as $SU(2)_L$ singlets~\cite{Silveira:1985rk,Pietroni:1992in,McDonald:1993ex}, doublets~\cite{Lee:1973iz,Haber:1984rc,Kim:1986ax,Peccei:1977hh,Turok:1990zg} and triplets~\cite{Konetschny:1977bn,Cheng:1980qt,Lazarides:1980nt,Schechter:1980gr,Magg:1980ut,Mohapatra:1980yp}. Furthermore, also models with an even more complex scalar sector, which are especially relevant for the excesses discussed in this work, such as the next-to-minimal two-Higgs-doublet model (N2HDM)~\cite{He:2008qm,Grzadkowski:2009iz,Logan:2010nw,Boucenna:2011hy,He:2011gc,Bai:2012nv,He:2013suk,Cai:2013zga,Chen:2013jvg,Guo:2014bha,Wang:2014elb,Drozd:2014yla,Ko:2014uka,Campbell:2015fra,Drozd:2015gda,vonBuddenbrock:2016rmr,Arhrib:2018qmw,Engeln:2020fld,Azevedo:2021ylf,Biekotter:2022abc,Banik:2023ecr,Biekotter:2023oen}, are frequently studied. 

On the experimental side, LHC searches for new particles in general, and new scalars in particular, have not led to any discovery yet. However, the searches for {additional} Higgses have been mostly performed inclusively or with a limited number of topologies, such that {significant} regions of the phase-space remain unexplored. In particular, associated production received relatively little attention. In this context, in recent years the so-called ``multi-lepton anomalies" emerged, which {are constituted by several} tensions in channels with multiple electrons and/or muons in the final states. These {discrepancies} are statistically {significant} and point towards associated production of EW scale new scalars~\cite{vonBuddenbrock:2016rmr,vonBuddenbrock:2017gvy,vonBuddenbrock:2019ajh,vonBuddenbrock:2020ter,Hernandez:2019geu,Fischer:2021sqw}. {In particular,} the multi-lepton anomalies are compatible with the direct production of a scalar $H$, with a mass of $\approx$270\,GeV, that decays dominantly into a pair of lighter scalars, $S$. A sub-set of these anomalies contains non-resonant {opposite sign, different flavour} di-leptons final states (with and without the presence of $b$-quark jets), pointing towards the decay $S\rightarrow W^+W^-\rightarrow\ell^+\ell^-, \ell=e,\mu$ with $m_S=150\,\pm5$\,GeV~\cite{vonBuddenbrock:2017gvy}. 

In Ref.~\cite{Crivellin:2021ubm}, we showed that the side-bands of the SM Higgs boson analyses of ATLAS~\cite{ATLAS:2020pvn,Aad:2020ivc,Aad:2021qks} and CMS~\cite{Sirunyan:2021ybb,Sirunyan:2020sum,CMS:2018nlv,Sirunyan:2018tbk} in fact suggest the presence of a {narrow scalar resonance} with a mass of $\approx\!151\,$GeV, produced in association with leptons and {($b-$)}jets.  
Furthermore, several hints for the existence of a new neutral scalar $S^\prime$ with a mass of $\approx\!95\,$GeV were presented by the CMS experiment~\cite{CMS:2018cyk,CMS:2023yay,CMS:2022goy}. {While previous ATLAS analyses did not exclude this potential signal~\cite{ATLAS:2018xad,ATLAS:2022yrq},} the latest {result}~\cite{ATLAS:2023jzc} shows a weaker-than-expected limit at this mass. Furthermore, an old LEP measurement suggests $e^+e^-\to Z^*\to S^\prime Z$ with $Z\to bb$~\cite{LEPWorkingGroupforHiggsbosonsearches:2003ing} and CMS finds a hint for resonant $\tau$ pair production at a similar mass~\cite{CMS:2022goy}. 

In this article, we combine, for the first time, the hints for a 95$\,$GeV scalar and extend and update the fit of Ref.~\cite{Crivellin:2021ubm} {for the $\approx\!151\,$GeV one} by including the {side-bands of the} recently released analysis of {associate production of the SM Higgs~\cite{ATLAS:2023omk}} as well as the $WW$ channel analyzed in Ref.~\cite{Coloretti:2023wng}, in order to obtain combined evidence for new scalar resonances at the EW scale.

\section{Channels}

\subsection{Low mass range: $S^\prime$ ($\approx95$\,GeV)}

$\boldsymbol{(S^\prime\to bb)+Z}$: LEP reported an excess with a local significance of 2.3$\sigma$ at $\approx98$\,GeV in Higgsstrahlung, i.e.~$e^+e^-\to ZS^\prime$ with $S^\prime\to bb$~\cite{LEPWorkingGroupforHiggsbosonsearches:2003ing}. We will not include this excess directly in the combination, but rather use it to obtain a more narrow mass window of $93\,{\rm GeV}<m_{S^\prime}<103\,$GeV to reduce the look-elsewhere-effect (i.e.~the trail factor). 

$\boldsymbol{S^\prime \to \gamma \gamma}$: We use the $p$-value graph in Fig.~7 of the CMS analysis~\cite{CMS:2023yay} and Figure~7~(a) of the ATLAS analysis~\cite{ATLAS:2023jzc}. 

$\boldsymbol{{S^\prime \to \tau \tau}}$: The relevant $p$-value graph is provided in the supplemental material of Ref.~\cite{CMS:2022goy} and shows an excess  with a local significance of $3.1\sigma$. While ATLAS did not perform an explicit search for new scalars in this final state, Ref.~\cite{ATLAS:2022yrq} see no excess in the side-band of the SM Higgs boson analysis. We therefore treat this as a null result which reduces the significance of the CMS excess by a factor of $\sqrt{2}$, assuming that the ATLAS and CMS analyses have similar sensitivity.

$\boldsymbol{{S^\prime \to W W^{*}}}$: We use the combined transverse mass distributions of lower graphs of Fig.~2 in Ref.~\cite{CMS:2022uhn} (CMS) and the upper-left graph of Fig.~11 in Ref.~\cite{ATLAS:2022ooq} (ATLAS). The details of the combination and the simulation are described in Ref.~\cite{Coloretti:2023wng} where an excess with a local significance of $\approx2.6\sigma$ was found.

\subsection{High mass range: $S$ ($\approx152$\,GeV)}

We utilise CMS and ATLAS studies of SM Higgses, which essentially encompass the search for other resonances in their side-bands. Depending on the channel, they range up to 180$\,$GeV. However, because some analyses stop at 160$\,$GeV. Since we want to avoid to be too close to the SM Higgs resonance, we will utilise the region between 140$\,$GeV and 155-160$\,$GeV, when appropriate.\footnote{Importantly, this mass range is suggested by, and compatible with, the multi-lepton anomalies.}

\begin{figure*}[t]
\centering
    \includegraphics[width=0.7\linewidth]{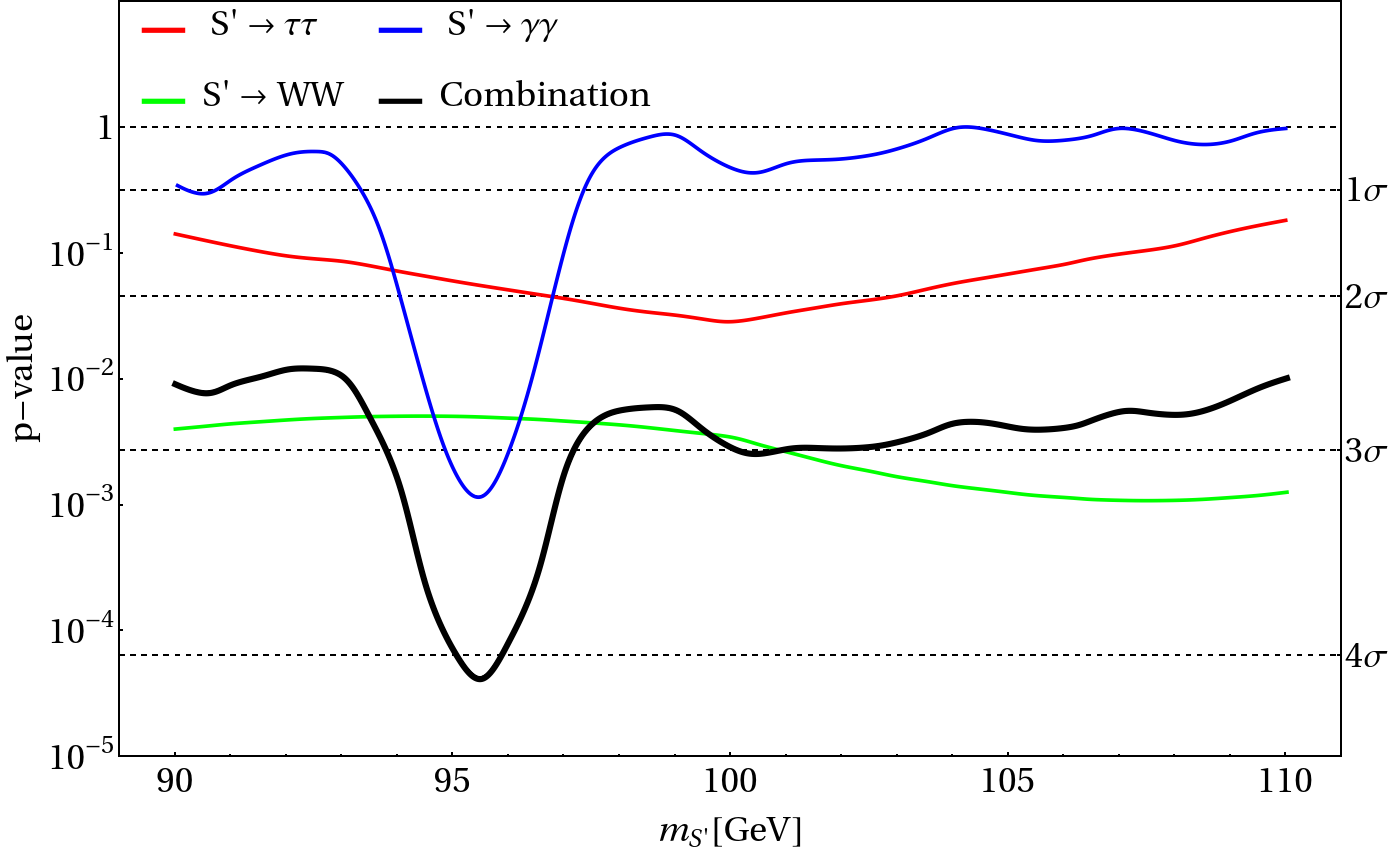}
\caption{The $p$-value as a function of the mass of $S^\prime$ for the low-mass channels (see main text for details).}\label{fig:Pvalue95All}
\end{figure*}

The combination will be performed in two steps. The first combination includes data reported up until 2021.

$\boldsymbol{{(S\to Z(\to \ell^+ \ell^-)\gamma) + \ell }}$: The invariant mass of the $Z\gamma$ pair is used to reconstruct $m_S$ and, in addition, the presence of an extra lepton (apart from the leptons produced by the $Z$ decay) is required (Fig.~5 in Ref.~\cite{CMS:2018myz}). 

$\boldsymbol{{(S\to \gamma \gamma) + E^T_{\rm miss} }}$: $m_S$ is reconstructed from the invariant mass of the photon pair and moderate additional missing energy of $\approx 100$\,GeV is required\footnote{The range of missing energy is dictated by a predefined simplified model of $H\rightarrow SS^{*}$ and it is not determined by the experimental analysis.} (see Fig.~6 in Ref.~\cite{ATLAS:2021jbf} and Fig.~3 in Ref.~\cite{CMS:2018nlv}). 

$\boldsymbol{{(S\to b \bar{b}) + E^T_{\rm miss} }}$: $m_S$ is reconstructed from the invariant mass of the bottom quark pair and missing transverse energy is required (Fig.~14 (a) in Ref.~\cite{ATLAS:2020fcp}). Here, some overlap of this category and $t\bar{t}h$ production in the SM exists. However, the contamination of the former by the latter due to the production of $E^T_{\rm miss}$ from the decay of $W$s is quite minimal (approximately 1$\%$).

$\boldsymbol{{(S\to \gamma \gamma) + b}}${\bf{-jet}}: In this channel, $S$ decays to two photons and is produced in association with at least a $b$-jet. The experimental data is extracted from Fig.~2 (top-right) in Ref.~\cite{ATLAS:2020ior} and Fig.~2 in Ref.~\cite{CMS:2020cga}.

$\boldsymbol{{(S\to \gamma \gamma) + W,Z }}$: In this study, $S$ decays to two photons and is produced in association with a weak gauge boson ($W$ or $Z$). The relevant data is obtained from Fig.~15 (bottom-left) in Ref.~\cite{CMS:2021kom} and Fig.~9 c) and d) in Ref.~\cite{ATLAS:2020pvn}. 

$\boldsymbol{{S\to \gamma \gamma}}$ {\bf (inclusive)}: Also in the (quasi-)inclusive case, $m_S$ is reconstructed from the invariant mass of the photon pair. However, vector boson fusion, as well as the presence of additional $W$ and $Z$ bosons and top quark-associated production are not included. Note that even though there is no veto on missing energy, the $S(\to \gamma \gamma) + E^T_{\rm miss}$ channel only covers a small portion of phase space of the quasi-inclusive final search. Here we use Fig.~15 (top-left) of Ref.~\cite{CMS:2021kom} and Fig.~9 a) of Ref.~\cite{ATLAS:2020pvn}.

This determines a very narrow mass range of interest, thus avoiding further scanning, and removing look-else-where effects when including new data. In the second step, data from the following recent CMS and ATLAS analysis will be added to the first combination:

$\boldsymbol{{(S\to \gamma \gamma)}+ \geq 4j}$: Here $m_S$ corresponds to invariant mass of di-photon pair which is produced in association with at least 4 jets (Fig.~2 a) in Ref.~\cite{ATLAS:2023omk}). 

$\boldsymbol{{(S\to WW^{(*)}}) + E^T_{\rm miss}}$: The CMS and ATLAS analyses of the SM Higgs boson decaying to a pair of $W$ bosons are recast and combined. Here we use the 0-jet category for which the dominant contribution from the simplified model described above arises from $H\rightarrow S(\rightarrow WW)S^{*}(\rightarrow E^T_{\rm miss})$. Other final states from associated production have very small jet veto survival probability. For ATLAS, we have used the data from Fig.~11 of Ref.~\cite{ATLAS:2022ooq} and for CMS the $m_T$ distributions ($p_{T2} < 20 \,$GeV and $p_{T2} > 20\,$GeV) of Fig.~1 of Ref.~\cite{CMS:2022uhn}. 

$\boldsymbol{{S(\to \gamma \gamma)}+}\boldsymbol{ \geq 1  \ell+b-}${\bf jet}: $S$ decays to two photons and is produced in association with at least one electron or muon ($\ell$) and at least one tagged $b$-jet. The relevant experimental data are taken from Fig.~5 a) in Ref.~\cite{ATLAS:2023omk}.

$\boldsymbol{S(\to \gamma \gamma) + \gamma}$: $S$, whose mass is obtained from the invariant mass of the leading and sub-leading photon, is produced in association with at least one additional photon moderate by $p_T \geq 25\,$GeV (see Fig.~6 a in Ref.~\cite{ATLAS:2023omk}).

$\boldsymbol{{S \to e \mu}}$: Here we combine the data from ATLAS and CMS using the right-hand panel of Fig.~1 in Ref.~\cite{ATLAS:2019old} and Fig.~8 of Ref.~\cite{CMS:2023pqk}. However, since this signal is exotic and not easy to account for in a UV complete model~\cite{Primulando:2023ugc,Afik:2023vyl}, we will show both the combinations with and without including this channel.

\section{Combinations}

\subsection{Low mass range (95 GeV)}

First, we combine the $\gamma\gamma$ results from ATLAS and CMS by assuming that both experiments have the same sensitivity to the signal see blue line in Fig.~\ref{fig:Pvalue95All}. We then add to this the $\tau\tau$ and $S \to W W^{*}$ signals, using Fisher's combined probability test~\cite{fisher1925} with three degrees of freedom (DOF):
\begin{equation}
\chi_{2 n}^2=-2 \sum_{i=1}^n \log \left(p_i\right),
\label{eq:Sig}
\end{equation}
where $p_i$ is the $p$-value of each channel in the combination and $\chi_{2 n}^2$ represents the chi-squared distribution with $2n$ degrees of freedom, where $n$ is the number of channels being combined. 

The resulting $\chi^2$ distribution is used to calculate the combined $p$-value shown in Fig.~\ref{fig:Pvalue95All}. The highest local significance of $4.1\sigma$ is obtained at $m_{S^\prime}\approx 95$\,GeV. Taking into account the LEP excess which narrows the mass range, the look-elsewhere effect, for a trial factor~\cite{Gross:2010qma,Barlow:2019svl} of $5/1.5\approx 3.3$ (obtained from dividing half the mass range by the resolution of 1.5$\,$GeV)~\cite{Gross:2010qma}, results in a global significance of $3.8\sigma$.

\begin{figure*}[t]
\centering
    \includegraphics[width=0.75\linewidth]{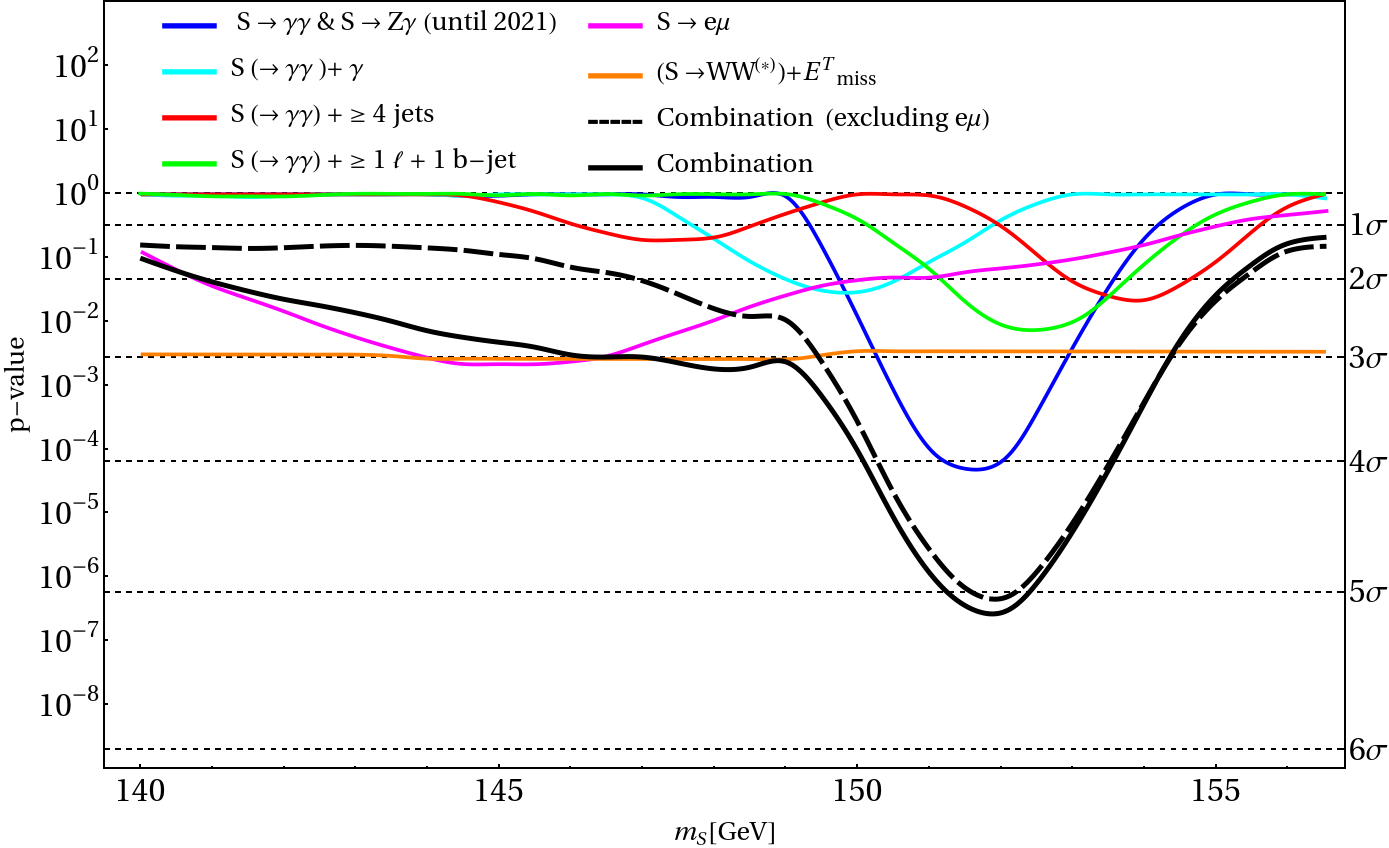}
\caption{The $p$-values of the individual high mass channels as well as their combination, both including and excluding the $\mu e$ signal.}\label{fig:Pvalue152All}
\end{figure*}

\subsection{High mass range (152 GeV)}

The required main production mechanism here is clearly associated production. Therefore, we assume a simplified model with a scalar $H$, with a mass of $270\,$GeV and directly produced via gluon fusion (as motivated by the multi-lepton anomalies), which decays into two lighter ones where one of them is off-shell ($SS^*$).\footnote{Refs.~\cite{vonBuddenbrock:2016rmr,vonBuddenbrock:2018xar} considered $H\rightarrow Sh$. While this has similar signatures than $H\to SS^*$, it leads to problems with SM Higgs signal strength measurements~\cite{CMS:2014nkk,ATLAS:2015zhl,CMS:2014fzn,ATLAS:2015egz,CMS:2022dwd,ATLAS:2022vkf} if it is the dominant decay mode.} We assume $S$ to be SM-like, i.e.~to have the branching ratios of a hypothetical SM Higgs with the same mass~\cite{LHCHiggsCrossSectionWorkingGroup:2016ypw,Braaten:1980yq,Sakai:1980fa,Inami:1980qp,Gorishnii:1983cu,Gorishnii:1990kd,Gorishnii:1990zu,Gorishnii:1991zr, Djouadi:1997rj, Degrassi:2005mc,Passarino:2007fp,Actis:2008ts,Spira:1991tj}.

In the first step, we assume an additional branching ratio to invisible final states. This means that the combination is thus performed with two DOFs. Considering only $\gamma\gamma$ and $Z\gamma$ channels for the on-shell $S$, 
we combine that data reported up to 2021.\footnote{The observed yields of the $\gamma\gamma$ and $Z\gamma$ are consistent with those predicted by the simplified model. Other channels are considered in the second combination pass.} The blue line in Fig.~\ref{fig:Pvalue152All} is obtained, corresponding to a maximal local significance of 4.0$\sigma$ at 152\,GeV.\footnote{The largest local significance for this first combination is $4.1\sigma$ at 151.5\,GeV. However, the largest significance in the global combination is obtained for 152\,GeV. As such, the significance is reported for this mass.} The trials factor is computed taking into account the different signal resolutions and the mass range (140-155\,GeV). This reduces the significance to 3.8$\sigma$. This result is combined with the results of the second combination for 152\,GeV.

In order to verify the consistency between the observed signal yields of the data released after 2021 within this simplified model, we simulated the processes, $p p \to H \to S S^* $, where $S$  decays to $S \to \gamma\gamma(Z\gamma)$ and $S^*$ decays to $b \bar{b}$, $ \tau^+ \tau^- $, $W W^*$ and missing energy. Again, we assumed that the ratios of the branching fractions of $b \bar{b}$ vs $ \tau^+ \tau^- $ and $W W^*$ are SM-like. Afterwards, we applied the event selection criteria detailed in Ref.~\cite{ATLAS:2023omk} to extract the signal efficiency of ${{S(\to \gamma \gamma)}}$ + ${ \geq 4}$ {{jets}}. The resulting expected yields from the simulation are found to be in agreement within $1\sigma$ with the observed yields from ATLAS. 

A similar procedure is followed to analyze the cross-section for ${{S(\to WW^{(*)}})  + E^T_{\rm miss}}$. The ratio of the extracted cross-sections of ${{S(\to \gamma \gamma) + E^T_{\rm miss} }}$ to that of ${{S(\to WW^{(*)}})  + E^T_{\rm miss}}$ is also consistent with the prediction of the simplified model. However, for purely SM-like branching ratios, the simplified model predicts an excess in $S\rightarrow ZZ^{*}\rightarrow 4\ell$, which is not  observed. As such, the significance of the ${{S(\to WW^{(*)}}) + E^T_{\rm miss}}$ is included in the combination with an additional DOF.

Because ${(S\to \gamma \gamma) \geq 1 \ell}+b$-jet and $(S\to \gamma \gamma) + \gamma$ are not predicted by the simplified model, they are added using two additional DOF. The invariant mass spectra of the channels being combined are fitted with the sum of background and signal functions described in 
Eq.~(\ref{background}) and Eq.~(\ref{eq:DSCB}). The parameterization of each function takes into account the signal resolution of the channel, while the background corresponds to the SM hypothesis.Figure~\ref{fig:Pvalue152All} shows the local $p$-value for the considered channels separately, where the significance for $(S\to \gamma \gamma) + \gamma$, $(S\to \gamma \gamma) + \geq 4$\,jets and $(S\to \gamma \gamma)+ \geq \ell+b$-jet are calculated individually using the formula~\ref{eq:poissonSig} with the weighted signal efficiency $\epsilon$ is equal to one. 

We proceed similarly with the $S \to e \mu$ channel, which is not present in the simplified model. The combined results from ATLAS and CMS of $S \to e \mu$ channel are detailed in Appendix~\ref{fig:emu}. Finally, all channels are combined using Fisher's combined probability Eq.~\ref{eq:Sig} with five DOFs. Figure~\ref{fig:Pvalue152All} displays the results, where a global significance of $5.0\sigma$ is found for $m_S=152$\,GeV. Since the $S\to e\mu$ signal is exotic, we also show the combination without this channel, leading to a global significance of $4.9\sigma$.

\section{Conclusions}
The multi-lepton anomalies signify the current statistically most significant deviation of LHC data from the SM predictions. They can be consistently explained by assuming a simplified model in which a heavy scalar $H$ decays into two lighter scalars $S$ with EW scale masses. Assuming a sizable decay width for $S\rightarrow W^+W^-\to\ell^+\ell^-,\ell=e,\mu$, the mass of the scalar was determined to be $m_S=150\,\pm5$\,GeV. 

Motivated by these anomalies and their possible explanations, we searched for narrow resonances in the side-bands of SM Higgs analysis and found a hint for a $\approx 151$\,GeV scalar~\cite{Crivellin:2021ubm}. In this article, we added to this combination the recent ATLAS and CMS analyses released after 2021 and found that the significance is further strengthened: assuming a simplified model with 5 DOF, we found $\approx\!5\sigma$ for $m_S\approx 152\,$GeV.

Furthermore, we combined the hints for the presence of a $\approx\!95\,$GeV scalar $S^\prime$, finding a preference of $3.8\sigma$ over the Standard Model hypothesis. This opens the possibility of a decay chain $H\rightarrow SS^{\prime}$ explaining the multi-lepton anomalies. In this case, the decay of the $S$ would be the source of leptons, while $S^{\prime}$ would be the origin of $b$-quarks. Finally, the absence of a $S\to ZZ^*\to 4\ell$ signal suggests that $S$ could be the neutral component of an $SU(2)$ triplet~\cite{Chabab:2018ert,FileviezPerez:2022lxp, Cheng:2022hbo,Chen:2022ocr,Rizzo:2022jti,Chao:2022blc,Wang:2022dte,Shimizu:2023rvi,Lazarides:2022spe,Senjanovic:2022zwy,Crivellin:2023gtf,Chen:2023ins,Ashanujjaman:2023etj,Ashanujjaman:2023etj}, as motivated by average~\cite{deBlas:2022hdk} of the $W$ mass measurement~\cite{CDF:2022hxs,ATLAS:2023fsi,LHCb:2021bjt,ALEPH:2013dgf}. 

\begin{figure*}[htbp]
\centering
    \includegraphics[width=0.7\linewidth]{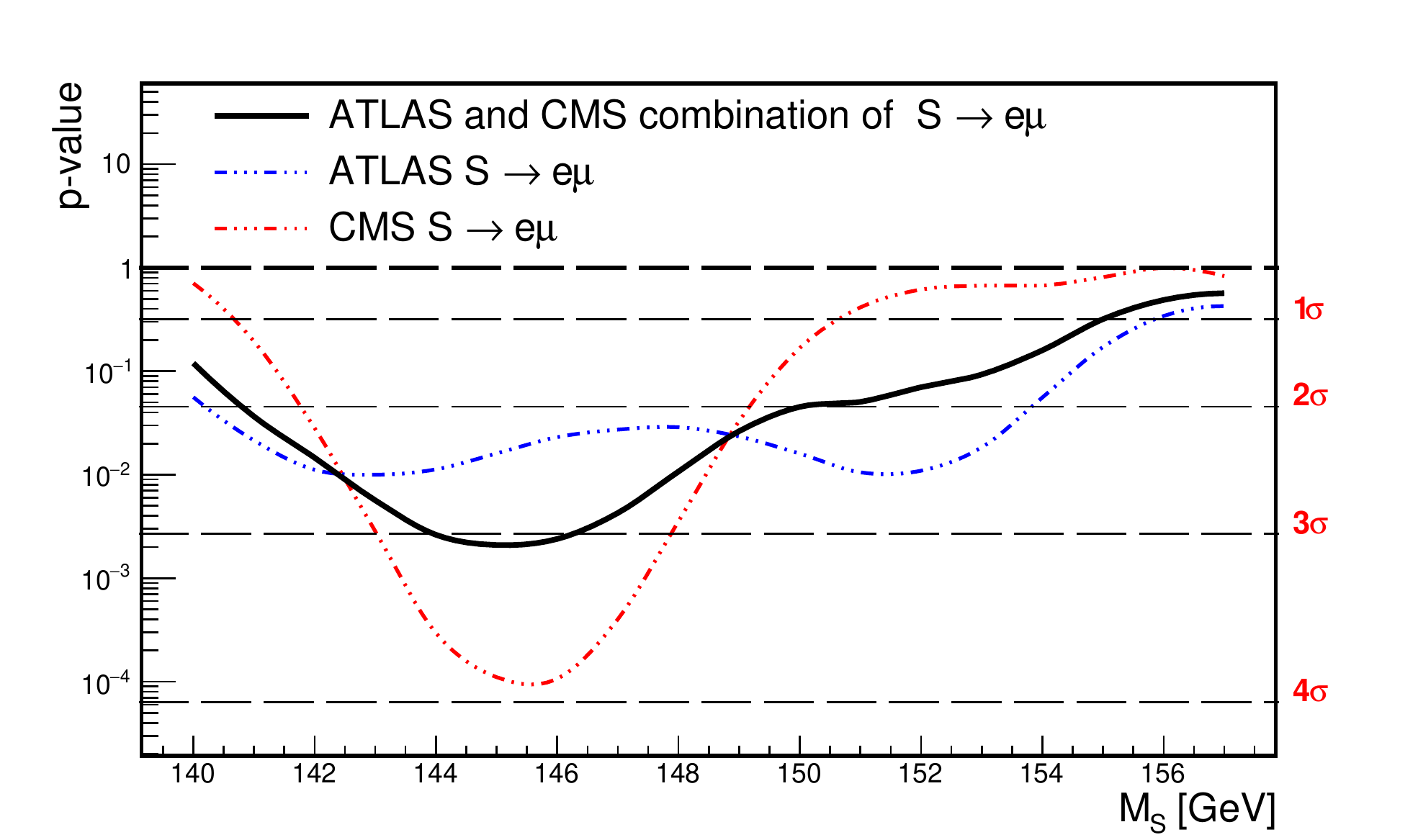}
    \label{fig:emu}
\caption{Individual and combined $p$-values for $S^\prime\to e\mu$ as a function of the mass.}
\end{figure*}

\begin{acknowledgments}
The work of A.C.~is supported by a professorship grant from the Swiss National Science Foundation (No.\ PP00P21\_76884). B.M.~gratefully acknowledges the South African Department of Science and Innovation through the SA-CERN program, the National Research Foundation, and the Research Office of the University of the Witwatersrand for various forms of support.
\end{acknowledgments}

\appendix

\section{Details of the analysis}
\subsection{$S$ $\rightarrow$ $ \gamma\gamma $}
As outlined in the introduction, we use CMS and ATLAS analyses to search for new scalars in the mass range between 90$\,$GeV  and 110$\,$GeV as well as between 140$\,$GeV and $155\,$GeV. For the side-band searches, for each category, we model the background via
\begin{equation}
f(m;b,\{a\}) = (1 - m)^b (m)^{a_0 + a_1 \log(m)}\,,
\label{background}
\end{equation}
where $a_{0,1}$ and $b$ are free parameters (different for each category) and $m$ is the invariant mass of the distribution, e.g.~the di-photon mass. The choice of the functional form to model the background is not important for our study, as shown in Ref.~\cite{Crivellin:2021ubm} to which the reader is refereed to for more technical details.

We add to the background parametrized by Eq.~(\ref{background}) a double-sided-crystal-ball function:
\begin{equation}
   N\cdot %\left\{
   \begin{cases}
   e^{-t^{2}/2} & \mbox{if $-\alpha_{\rm Low} \leq t \leq \alpha_{\rm High}$}\\
   \frac{ e^{-0.5\alpha_{\rm Low}^{2} }}{ \left[\frac{\alpha_{\rm Low}}{n_{\rm Low}} \left(\frac{n_{\rm Low}}{\alpha_{\rm Low}} - \alpha_{\rm Low} -t \right)\right]^{n_{\rm Low}} } & \mbox {if $t < -\alpha_{\rm Low} $}\\
   \frac{ e^{-0.5\alpha_{\rm High}^{2} }}{ \left[\frac{\alpha_{High}}{n_{\rm High}} \left(\frac{n_{\rm High}}{\alpha_{\rm High}} - \alpha_{\rm High}  + t \right)\right]^{n_{\rm High}} } & \mbox {if $t > \alpha_{\rm High}$}.
   \end{cases}
   \label{eq:DSCB}
\end{equation}
Here $N$ is a normalization parameter, $t = (m - m_{S})/\sigma_{CB}$ where $\sigma_{CB}$ is the width of the Gaussian part of the function, $m$ is the invariant mass of the distribution and $m_S$ the mass of the new scalar. 

\subsection{$S \rightarrow WW$}
We have simulated the process $pp \rightarrow H \rightarrow S S^* ,S \rightarrow W^+W^{-(*)} \rightarrow \ell^+ \ell^- \nu \bar{\nu}$ and $S^*$ going to missing energy. Note that this is dominant compared to $S^*\to WW$ and $S$ going to missing energy if the mass of the invisible particle is small. To validate and improve our fast simulation we simulated the SM Higgs boson signal, i.e. $gg \rightarrow h \rightarrow W W^{(*)} \rightarrow \ell^+ \ell ^- \nu \bar{\nu}$ and compared to the ATLAS one for the SM Higgs boson signal given as a function of the transverse mass $m_T$ in Fig.~11 in Ref.~\cite{ATLAS:2022ooq}. In addition, smearing etc.~was applied to correct for our fast simulation as explained in Ref.~\cite{Coloretti:2023wng}.

%\subsection{Simplified Model}
%In order to verify or falsify the hypothesis of sizable $S\to b\overline{b}$ or $H\to Sh$ rates, one could search for $H\rightarrow SS\rightarrow\gamma\gamma b\overline{b}$ and $H\rightarrow Sh\rightarrow\gamma\gamma b\overline{b}$ final states. These are very promising signatures as they have the highest sensitivity for di-Higgs searches due to a good balance between the di-photon triggering efficiency, the triggering of the invariant mass spectra~\cite{CMS:2018tla,ATLAS:2018dpp}.

%We illustrate this for $H\rightarrow SS\rightarrow\gamma\gamma b\overline{b}$. Assuming $m_H=270\,$GeV, the dominant branching ratio being $H\to SS^*$ forces one of the singlet scalars to be off-shell. This type of resonant $\gamma\gamma b\overline{b}$ searches have not been performed by the LHC experiments. Here, two corners of the phase-space are devised to study asymmetric configurations: $m_{\gamma\gamma}\in(145,155)$\,GeV and $m_{b\overline{b}}\in(70,120)$\,GeV to isolate $H\rightarrow S(\rightarrow \gamma\gamma) S^*(\rightarrow b\overline{b})$; $m_{\gamma\gamma}\in(90,120)$\,GeV and $m_{b\overline{b}}\in(120,160)$\,GeV to isolate $H\rightarrow S(\rightarrow b\overline{b}) S^*(\rightarrow\gamma\gamma)$. 

\subsection{Combination of ATLAS and CMS data of the $ S \rightarrow e \mu $ channel}

The combination of ATLAS~\cite{ATLAS:2019old} and CMS~\cite{CMS:2023pqk} data are obtained from a simultaneous fit to the invariant mass of the electron–muon pair, in the mass range between $140$\,GeV and $157$\,GeV. The backgrounds in both ATLAS and CMS are parameterized using functional form in Eq.~(\ref{background}) with different normalization factor to take in the account the difference in background contamination from each experiment. Afterward, we scanned the invariant mass spectrum of ATLAS and CMS simultaneously by adding a double-sided-crystal-ball function described in Eq.~(\ref{eq:DSCB}), the parameter $N$ of the added DSCB functions is considered as a common and free parameter, while the remaining parameters are fixed to the values extracted from the signal fit to Higgs-like scalar decaying to one electron and one muon in ATLAS and CMS detectors. Figure~\ref{fig:emu} shows the combination of ATLAS and CMS data in the $S \to e \mu $ decay channel, where the individual significance is estimated using the median significance formula~\cite{Barlow:2019svl}
\begin{equation}
S=\sqrt{2}\sqrt{\left(\epsilon \cdot S+B\right) \log \left(1+{\epsilon\cdot S}/{B}\right)-\epsilon \cdot S}.
   \label{eq:poissonSig}
\end{equation}
Here, $B$ and $S$ are the continuum background and signal yields respectively, which are extracted from the simultaneous fit. While, the weighted signal efficiencies $\epsilon$ are found to be $55\%$ and $45\%$, accordingly with the individual efficiency of the $S \to e \mu $ channel in the ATLAS and CMS analyses, respectively. Finally, the combined significance is calculated by summing the two individual significance in quadrature. After the combination, the ATLAS significance of $3.8\sigma$ at $m_S\approx146$\,GeV is reduced to $3.1\sigma$ by included the CMS result. Note that a significance of $\approx2\sigma$ around $m_S=151$\,GeV is found.

\bibliographystyle{apsrev4-1}
%\bibliographystyle{unsrt}

%\bibliography{apssamp}
\bibliography{apssampMukesh}

\end{document}